\documentclass[proceedings, preprint]{rmaa}

\usepackage{times}
\usepackage{graphicx}
\usepackage{ulem}
\usepackage{epsfig}
\usepackage{natbib}

\newcommand{\msun}{\mbox{$M_\odot$}}

\def\be{\begin{eqnarray}}
\def\ee{\end{eqnarray}}
\def\bi{\begin{itemize}}
\def\ei{\end{itemize}}
\def\lsim{\mathrel{\rlap{\lower3pt\hbox{\hskip1pt$\sim$}}
     \raise1pt\hbox{$<$}}} 
\def\gsim{\mathrel{\rlap{\lower3pt\hbox{\hskip1pt$\sim$}}
     \raise1pt\hbox{$>$}}} 

\SetYear{2013}
\SetConfTitle{Magnetic Fields in the Universe 4}

\title{Thermal Photons From Magnetized Bare Strange Stars} 

\author{Enrique Moreno M\'endez\altaffilmark{1}, Dany Page\altaffilmark{1}, Leonardo Pati\~no\altaffilmark{2}, Patricia Ortega\altaffilmark{2}}

\altaffiltext{1}{Instituto de Astronom\'ia, Universidad Nacional Aut\'onoma de M\'exico, Circuito Exterior, Ciudad Universitaria, \\Apartado Postal 70-543, 04510, Distrito Federal, M\'exico. (enriquemm, dany@astro.unam.mx).}

\altaffiltext{2}{Facultad de Ciencias, Universidad Nacional Aut\'onoma de M\'exico, Circuito Exterior, Ciudad Universitaria, \\Apartado Postal 70-543, 04510, Distrito Federal, M\'exico. (leopj@fciencias.unam.mx).}

\shortauthor{Moreno M\'endez et al.}
\shorttitle{RevMexAA(SC) Demo Document}

\listofauthors{Enrique Moreno M\'endez, Dany Page, Leonardo Pati\~no, Patricia Ortega}
\indexauthor{Moreno M\'endez, E., Page, D. Pati\~no, L., Ortega, P.}

\abstract{A plasma made out of strange-quark matter (SQM) and electrons, has a rather high plasma frequency ($\sim20$ MeV).  Thus, a compact star made of such material all the way up to its surface, i.e., a bare strange star, would be unable to radiate away its thermal emission.

We use the MIT-bag model and assume that SQM is the ground state of nuclear matter at high density.  We investigate whether the presence of a magnetic field will allow propagation of radiation at frequencies below the SQM plasma frequencies. Hence, we study the presence of gyrofrequencies in a SQM plasma permeated by a strong magnetic field ($B > 10^{12}$ G).  We find that small regions in the frequency spectrum allow radiation propagation due to the presence of the magnetic fields.  It is likely that narrow bands of radiation would likely be observable from magnetized bare strange stars .
}

\resumen{Como ha sido mostrado con anterioridad, la materia de quarks, y en particular, un plasma de materia de quarks extra\~nos y electrones tiene una frecuencia de plasma relativamente alta ($\sim20$ MeV).
Entonces, una estrella compacta compuesta de \'este material hasta su superficie, i.e., una estrella extra\~na desnuda, no podr\'ia enfriarse por emisi\'on de su radiaci\'on termica.

Usamos el modelo de la bolsa de MIT y suponemos que la materia extra\~na es el estado base de la materia nuclear a alta densidad.
Investigamos si la presencia de un campo magnetico permite la propagaci\'on de radiaci\'on a frecuencias inferiores a la del plasma de materia extra\~na. 
Esto es, estudiamos la presencia de girofrequencias en un plasma de materia extra\~na permeado por un campo magn\'etico fuerte ($B > 10^{12}$ G).
Encontramos que se abren peque\~nas ventanas de emisi\'on en el espectro de frecuencias dada la presencia de los campos magn\'eticos fuertes.
La longitud de onda de la radiaci\'on que se puede propagar depende linealmente con la magnitud del campo magn\'etico y de su orientaci\'on.
Esto indica que peque\~nas bandas de emisi\'on de radiaci\'on t\'ermica de estrellas extra\~nas desnudas podr\'ian ser observables.
}

\addkeyword{binaries: general}
\addkeyword{black hole physics}
\addkeyword{gamma-ray bursts: general}
\addkeyword{stars: evolution}
\addkeyword{stars: magnetic field}
\addkeyword{stars: massive}

\begin{document}
\maketitle

\section{Introduction}
\label{sec:Intro}

\begin{figure}[!t]
  \includegraphics[width=\columnwidth]{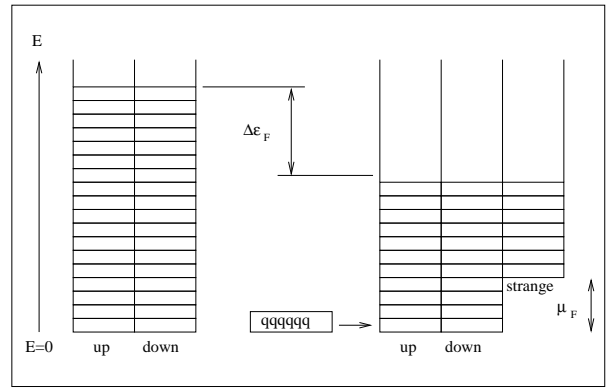}
  \caption{This diagram shows how the Fermi energy, $\epsilon_F$ of quark matter is lowered by a $\Delta \epsilon_F$ when the quarks are distributed in more flavors.  As can be seen, this may only happen when the chemical potential of the next lightest flavor is lower than the Fermi energy of the system.  Thus, no charm quarks are likely to appear in quark stars.}
  \label{fig:efermi}
\end{figure}

Stars with zero-age-main-sequence (ZAMS) masses $\gtrsim 8 \msun$ and less massive than some 20 to $25 \msun$ usually end their nuclear-burning lifetimes by having their cores collapsing into a neutron star (NS) and their envelopes blown away by the supernova (SN) launched by a shockwave created by the rebund of the homologous core and, likely, a revivival of the shock by neutrinos trapped behind it \citep[see, e.g.,][]{1979NuPhA.324..487B,1990RvMP...62..801B}.

The nuclear equation of state (EoS) is not known and, therefore, many EoSs have been proposed for cold, supranuclear matter \citep[among them, see, e.g., kaon condensates, by][]{1994ApJ...423..659B}.
With the recent measurements of NS masses around $2 \msun$ \citep{2010Natur.467.1081D,2013Sci...340..448A}, the minimum requirements on the stiffnes of the nuclear EoS have been further constrained \citep[see, e.g.,][]{2010arXiv1012.3208L}.
Among the proposed EoSs for matter denser than nuclear matter ($\rho_0\simeq10^{14}$ g cm$^{-3}$) is strange-quark matter (SQM).
The idea behind SQM is that as the hadron density grows a few times larger than nuclear density inside a neutron star the quarks may become de confined (thus forming quark matter, or QM).  
Now, according to \citet{Witten:1984rs}, the Fermi energy of this plasma of quarks can be lowered by converting roughly a third of the quarks (minus the chemical potential of the new quark species) to strange quarks (see~\ref{fig:efermi}).  
Thus creating SQM, and, hence, a strange star.
SQM may be fully stable if it is more bound than $^{56}_{26}$Fe, i.e., if it is the ground state of matter.
It must be noted that the original calculations for SQM had a rather difficult time explaining masses as high as $2 \msun$ \citep[see, e.g.,][]{2010arXiv1012.3208L}.  
However, \citet{2010PhRvD..81j5021K} have now performed O$(\alpha_c^2)$ estimates (where $\alpha_c$ is the strong coupling constant) of the EoS for SQM and find an upper limit around $2.75 \msun$ for the maximum mass of a strange star.

\begin{figure}[!t]
  \includegraphics[angle=270,width=\columnwidth]{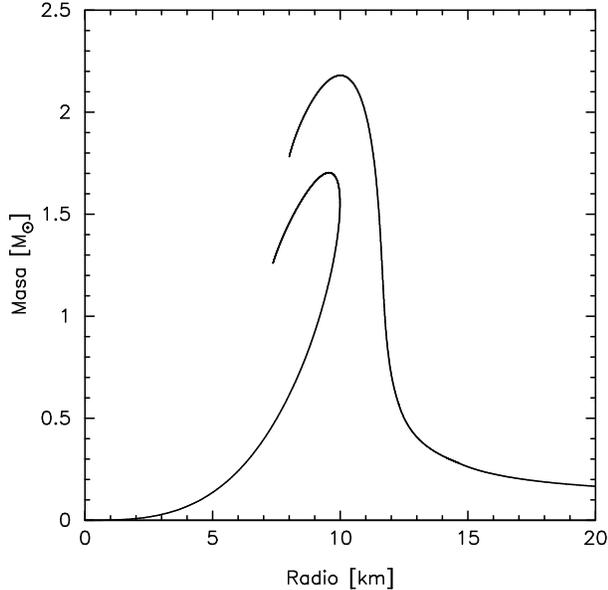}
  \caption{Plot of mass vs radius for a neutron star (righthand curve), and a strange star (left).}
  \label{fig:mr_str-APR}
\end{figure}

\begin{figure}[!t]
  \includegraphics[angle=270,width=\columnwidth]{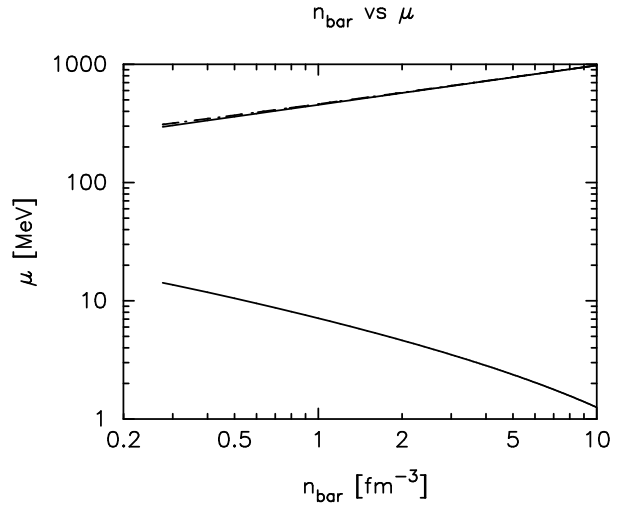}
  \caption{In this plot we observe how the chemical potential of each particle species varies with respect to the change in barion density.  It is clear the chemical potentials of each quark species is similar to the other ones, whereas that for electrons is over an order of magnitude smaller at the surface.}
  \label{fig:muvsnbar}
\end{figure}

We shall follow the procedure followed by \citet{1984PhRvD..30.2379F}, \citet{Haensel:1986qb},  \citet{1986ApJ...310..261A} and further explained by \citet{1997csnp.conf.....G}, i.e., we shall use the MIT bag model to find the chemical potentials of the different species in the quarks and electrons plasma.
The chemical equilibrium is maintained between the four species of particles in our strange matter model. Baryon number must be conserved as well.  Lastly, electric charge must be locally (as well as globally) neutral:

\begin{equation}
\mu_{s} = \mu_{d} \equiv \mu,
\label{eq:equim1}
\end{equation}
\begin{equation}
\mu_{e} + \mu_{u} \equiv \mu,
\label{eq:equim2}
\end{equation}
\begin{equation}
n_{bar} = \frac{1}{3}\left(n_{u} + n_{d} + n_{s} \right),
\label{eq:nbarcte}
\end{equation}
\begin{equation}
-n_{e} + \frac{2}{3}n_{u} - \frac{1}{3}n_{d} - \frac{1}{3}n_{s} = 0.
\label{eq:charge}
\end{equation}

\section{Strange Star Structure}
\label{sec:Struct}

\begin{figure}[!b]
  \includegraphics[angle=270,width=\columnwidth]{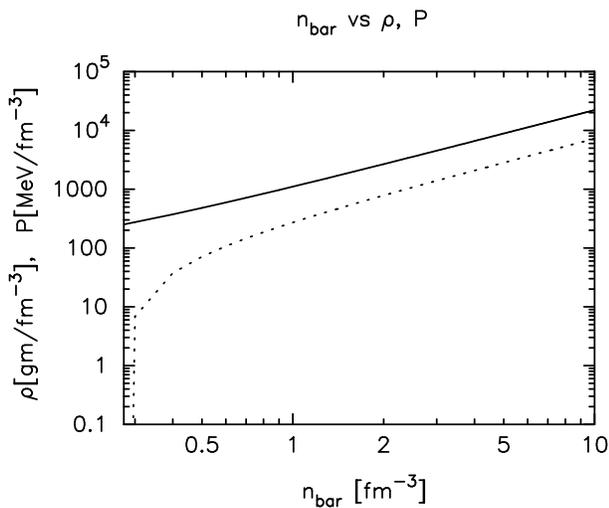}
  \caption{Presure (continous line) and density (dashed line) plotted versus barion density in our numerical model.}
  \label{fig:rhoypres}
\end{figure}

Using MIT’s bag model we obtain the structure of a compact star made out of SQM.  
Fig.~\ref{fig:mr_str-APR} shows the radius vs mass of a neutron star and a strange star.  
A plot of barion density vs the chemical potential of electrons, up, down and strange quarks can be seen on Fig.~\ref{fig:muvsnbar}.  
Finally, barion density vs density and pressure are displayed in Fig.~\ref{fig:rhoypres}.

\section{Gyrofrequencies in Magnetized SQM}
\label{sec:B-SQM}

\begin{figure}[!t]
  \includegraphics[angle=0,width=\columnwidth]{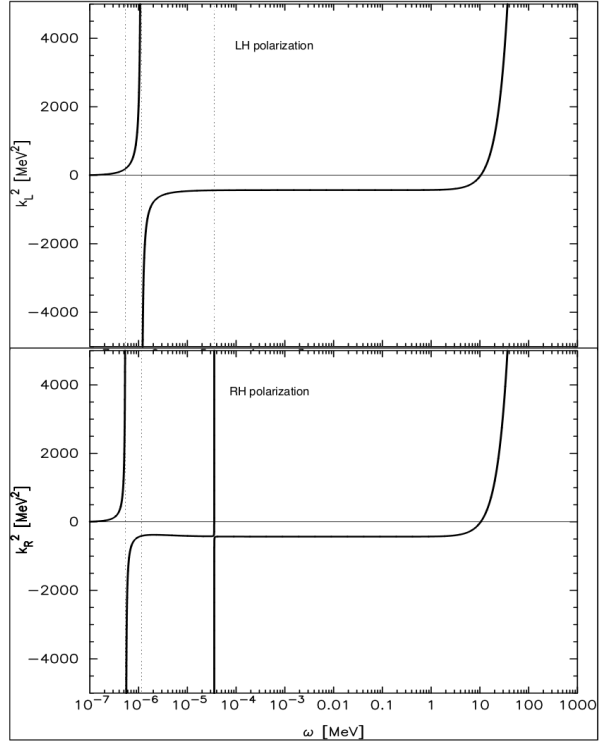}
  \caption{In these two plots we observe how the two polarization modes will have different emission bands below the plasma frequency.}
  \label{fig:k2_w-parLRing}
\end{figure}

\begin{figure}[!t]
  \includegraphics[angle=0,width=\columnwidth]{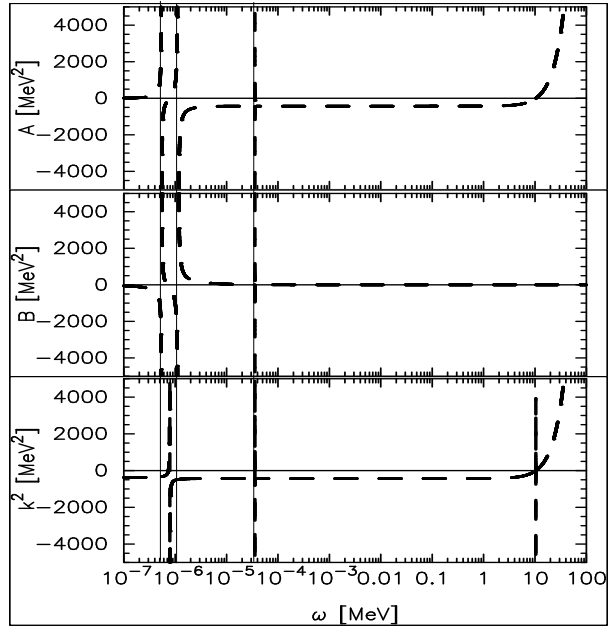}
  \caption{In this plots we observe how $A$ and $B$, two components of the matrix of the wave equation, give rise to the main properties of $k^2$.}
  \label{fig:k2vsw-perp}
\end{figure}

Using values for the strong coupling constant between $\alpha_C = 0$ and 0.9, the bag constant $B^{1/4} = 120$ and 165 MeV, the s quark mass ($100 < m_s < 200$ MeV$/c^2$) and a renormalization point \citep[chosen to be $\sigma \simeq 300$ as in][]{1984PhRvD..30.2379F}, we obtain plasma and gyrofrequencies.  We find the gyrofrequencies for EM waves propagating parallel (Fig.~\ref{fig:k2_w-parLRing}; left \& right polarization) and perpendicular (Fig.~\ref{fig:k2vsw-perp}) to the B field.

Further details will be presented in Moreno M\'endez \& Page (2013, in progress).

\section{AdS/CFT}
\label{sec:AdS}

Strictly speaking, quark matter, or the quark-gluon plasma (QGP), is a system that should be treated in the context of quantum field theory (QFT).  Unfortunately, this cannot be done in its perturbative formulation given its limitations regarding strong couplings.

The gauge/gravity correspondence \citep{1999IJTP...38.1113M} is a tool to carry this type of calculation that, even if not formally proved yet, has shown an overwhelming amount of evidence for its correctness and a relevant number of advantages over other methods.

In particular, the photon production by an isotropic QGP was computed in \citet{2007JHEP...11..025M} and recently extended to the anisotropic case \citep{013JHEP...02..154P}.

Following this lines, we can use the background constructed in \citet{2009JHEP...10..088D}, dual to a FT in the presence of a strong magnetic field, to compute both, the density of thermal photons produced by quark matter and its conductivity when a strong magnetic field is present.

This two are the key elements needed to take QFT in to account in the investigation presented in this poster.
Further details will be available in Pati\~no, Ortega, Moreno M\'endez et al. (2013, in progress).

\section{Conclusions}
\label{sec:Concl}

We have shown that both, the perpendicular and the parallel propagation modes for EM waves in a magnetized plasma of quarks and electrons find small windows where propagation is possible below the plasma frequency.  We have also shown that these frequency bands are due to the gyrofrequency of the particles that compose the magnetized plasma.

A strange star with a surface magnetic field of order $\sim10^{12}$ G may be able to emit EM radiation near the optical range, whereas a magnetar, where the surface magnetic field intensity is ~10e15 G, should emit near the X-ray band.  Bare strange stars have an outer layer of electrons which may further block part of this emission.

We are currently studying these effects with a newly available method, AdS/CFT (as described in window 4 below).  This method considers QFT and the details of the physical quantities needed for our analysis as well as other possible effects, which may be easily overlooked by methods such as the Bag model, like, e.g., superconductivity and/or superfluidity, and which may have important consequences.  

\end{document}